\documentclass[runningheads]{llncs}
\usepackage{tabularx}
\newcolumntype{Y}{>{\centering\arraybackslash}X}
\usepackage{hyperref}
\usepackage{graphicx}
\usepackage[misc,geometry]{ifsym}
\begin{document}
\title{Brain PET-MR attenuation correction with deep learning: method validation in adult and clinical paediatric data}
\titlerunning{Brain PET-MR attenuation correction with deep learning}
\author{Siti Nurbaya Yaakub\inst{1,2} \and
Colm J. McGinnity\inst{1} \and
Eric Kerfoot\inst{3} \and
Ines Mérida\inst{4} \and
Katherine Beck\inst{5} \and
Eric Dunston\inst{3} \and
Marica Muffoletto\inst{3} \and
Suryava Bhattacharya\inst{3} \and
Ahmed Qureshi\inst{3} \and
Joel Dunn\inst{1} \and
Jane MacKewn\inst{1} \and
Alexander Hammers\inst{1}\textsuperscript{\Letter}}
\authorrunning{S. N. Yaakub et al.}
\institute{King’s College London \& Guy’s and St Thomas’ PET Centre, School of Biomedical Engineering \& Imaging Sciences, King’s College London, London, U.K.\\
\email{alexander.hammers@kcl.ac.uk} \and
Brain Research \& Imaging Centre, University of Plymouth, Plymouth, U.K. \and
Department of Biomedical Engineering, School of Biomedical Engineering \& Imaging Sciences, King’s College London, London, U.K. \and
CERMEP-Imagerie du vivant, Lyon, France \and
Department of Psychosis Studies, Institute of Psychiatry, Psychology, and Neuroscience, King’s College London, London, U.K.}
\maketitle
\begin{abstract}
Current methods for magnetic resonance-based positron emission tomography attenuation correction (PET-MR AC) are time consuming, and less able than computed tomography (CT)-based AC methods to capture inter-individual variability and skull abnormalities. Deep learning methods have been proposed to produce pseudo-CT from MR images, but these methods have not yet been evaluated in large clinical cohorts. Methods trained on healthy adult data may not work in clinical cohorts where skull morphometry may be abnormal, or in paediatric data where skulls tend to be thinner and smaller. Here, we train a convolutional neural network based on the U-Net to produce pseudo-CT for PET-MR AC. We trained our network on a mixed cohort of healthy adults and patients undergoing clinical PET scans for neurology investigations. We show that our method was able to produce pseudo-CT with mean absolute errors (MAE) of 100.4 ± 21.3 HU compared to reference CT, with a Jaccard overlap coefficient of 0.73 ± 0.07 in the skull masks. Linear attenuation maps based on our pseudo-CT (relative MAE = 8.4 ± 2.1\%) were more accurate than those based on a well-performing multi-atlas-based AC method (relative MAE = 13.1 ± 1.5\%) when compared with CT-based linear attenuation maps. We refined the trained network in a clinical paediatric cohort. MAE improved from 174.7 ± 33.6 HU when using the existing network to 127.3 ± 39.9 HU after transfer learning in the paediatric dataset, thus showing that transfer learning can improve pseudo-CT accuracy in paediatric data.
\end{abstract}
\section{Introduction}
Brain positron emission tomography (PET) obtains images from radioactivity detected via 511 keV photons emerging inside the brain after positron emission. These photons undergo attenuation, especially in dense material like the skull, and attenuation correction (AC) is crucial for PET quantification. Standard PET coupled to computed tomography (CT) uses the density information from CT for AC (CT-AC). With the advent of hybrid (simultaneous) PET and magnetic resonance (MR) imaging, this is no longer possible, and MR-based methods (MR-AC) had to be developed. 

The current best-performing methods for brain MR-AC are multi-atlas co-registration methods, which give average PET quantification errors within ±3\% of CT-AC~\cite{Ladefoged2017}. However, these methods rely on multiple co-registrations with healthy control atlas sets of CT and MR images, and thus tend to be time consuming, and less able than CT to capture inter-individual variability and model abnormalities. Deep learning methods for synthesising pseudo-CT from MR images are emerging ~\cite{Chen2021,Klaser2021} but many are trained and tested on healthy data only, and very few have been evaluated in large clinical cohorts~\cite{Ladefoged2020} and in children~\cite{Ladefoged2019}. 

Children often require general anaesthesia for scanning. Simultaneous PET-MR allows scanning in a single session, avoiding repeat anaesthesia. However, children have smaller heads and thinner skulls than adults, so methods that perform well in adults may not necessarily perform well in children.

Here we aimed to show that transfer learning of a network trained on adult data can improve pseudo-CT accuracy in paediatric data. We trained our network on a cohort of primarily adult research data and evaluated our deep learning MR-AC against the CT-AC, used as the gold standard, and a well-performing~\cite{Ladefoged2017} multi-atlas-based AC method, MaxProb~\cite{Merida2017}. We then used a simple transfer learning approach, implementing and refining the network trained on adult data in a clinical paediatric cohort. We showed that pseudo-CT accuracy improved with transfer learning and that the refined network could produce pseudo-CT in paediatric data with accuracy comparable to the adult dataset.
\section{Materials}
\subsection{PET Centre Dataset}
The training dataset consisted of data from research studies at the KCL \& GSTT PET Centre (“PET Centre dataset”). Research studies were reviewed and approved by regional research ethics committees: West Midlands – Coventry \& Warwickshire Research Ethics Committee (16/WM/0364), and West London \& GTAC Research Ethics Committee (16/LO/0130). All participants provided written informed consent. 

This dataset consisted of 110 individuals who were scanned as part of three PET-MR research studies at the KCL \& GSTT PET Centre between 2016 and 2019 - the [\textsuperscript{18}F]fluorodeoxyglucose ([\textsuperscript{18}F]FDG) study, the [\textsuperscript{18}F]GE-179 study and the [\textsuperscript{18}F]Fallypride study. All low-dose reference CT data were acquired on a whole-body GE Discovery 710 PET/CT system on the same day as the MR scan. All MR data were acquired on a 3T Siemens Biograph mMR PET-MR system. Participant demographics and MR and CT acquisition details for each sub-group of data are given in Table 1. This dataset consisted primarily of adult data, however, there were two children aged 9 and nine teenagers between the ages of 13 to 17 years.
\begin{table}
\centering
\caption{PET Centre dataset: participant demographics and image acquisition details.} \label{table1}
\begin{tabularx}{0.9\textwidth}{ l *{4}{|Y} }
\hline
\multicolumn{2}{l|}{Tracer/Study} & [\textsuperscript{18}F]FDG & [\textsuperscript{18}F]GE-179 & [\textsuperscript{18}F]Fallypride\\
\hline
\multicolumn{2}{l|}{Number} & 57 & 31 & 22\\
\hline
\multicolumn{2}{l|}{Mean age ± SD} & 40.7 ± 19.2 & 24.3 ± 4.1 & 30.2 ± 8.7\\
\multicolumn{2}{l|}{(min–max) in years} & (9–79) & (18–36) & (18–55)\\
\hline
\multicolumn{2}{l|}{Female/Male} & 30/27 & 12/19 & 19/3\\
\hline
\multicolumn{1}{l|}{ } & Orientation & Sagittal & Sagittal & Sagittal\\
\multicolumn{1}{l|}{ } & TE (ms) & 2.63 & 2.63 & 2.19\\
\multicolumn{1}{l|}{ } & TR (ms) & 1.7 & 1.7 & 1.7\\
\multicolumn{1}{l|}{ } & TI (ms) & 900 & 900 & 900\\
MPRAGE & FA (°) & 9 & 9 & 9\\
parameters & Slice thickness (mm) & 1.1 & 1.1 & 1\\
\multicolumn{1}{l|}{ } & Num. slices & 176 & 176 & 240\\
\multicolumn{1}{l|}{ } & FoV (mm) & 270 & 270 & 256\\
\multicolumn{1}{l|}{ } & Percent phase FoV & 87.5 & 87.5 & 100\\
\multicolumn{1}{l|}{ } & Matrix size & 224 × 256 & 224 × 256 & 256 × 256\\
\multicolumn{1}{l|}{ } & Voxel size (mm) & 1.055 × 1.055 & 1.055 × 1.055 & 1 × 1\\
\hline
\multicolumn{1}{l|}{ } & FoV (cm) & 15 & 30 & 30\\
CT & kVp & 140 & 140 & 140\\
parameters & mAs & 8 & 8 & 8\\
\multicolumn{1}{l|}{ } & Voxel size (mm) & 0.5 × 0.5 × 2.5 & 0.5 × 0.5 × 2.5 & 0.5 × 0.5 × 2.5\\
\hline
\multicolumn{5}{p{.85\textwidth}}{TE: echo time; TR: repetition time; TI: inversion time; FA: flip angle; FoV: field of view.}\\
\end{tabularx}
\end{table}
\subsection{Clinical paediatric data}
The “clinical paediatric dataset” consisted of anonymised clinical scans acquired at Guy’s and St Thomas’ Hospital available through an NHS clinical service evaluation of patients with dystonia from 2013 to 2017 (Service Evaluation GSTT Reference 10319). Scans were performed as part of standard clinical assessment for possible deep brain stimulation and included MR imaging, [\textsuperscript{18}F]FDG-PET, and head CT under general anaesthesia. We included data from patients who had a complete set of [\textsuperscript{18}F]FDG-PET, 3D T1-weighted MR and CT scans, resulting in a dataset consisting of 49 patients between 2 and 17 years old (mean age ± SD = 9.3 ± 4.3 years). Patients underwent [\textsuperscript{18}F]FDG-PET-CT imaging on a GE (General Electric Medical Systems, Waukesha, WI) Discovery ST and a Discovery VCT scanner prior to October 2013, and thereafter on a GE Discovery 710 scanner, at the King’s College London \& Guy’s and St Thomas’ PET Centre. CT images had voxel sizes of 0.5 × 0.5 × 2.5 mm\textsuperscript{3} (n = 32) or 0.977 × 0.977 × 3 mm\textsuperscript{3} (n = 17). 3D T1-weighted images were acquired in sagittal orientation with voxel sizes of 0.694 × 0.694 × 0.68 mm\textsuperscript{3} (n = 15), 0.977 × 0.977 × 1 mm\textsuperscript{3} (n = 28), 0.446 × 0.446 × 1 mm\textsuperscript{3} (n = 1), 0.898 × 0.898 × 1 mm\textsuperscript{3} (n = 1), and approximately 1 × 1 × 1 mm\textsuperscript{3} (n = 2). One patient had their T1-weighted MR image acquired in transverse orientation (voxel size = 0.55 × 0.55 × 1.1 mm\textsuperscript{3}). 
\section{Methods}
\subsection{Data preprocessing}
T1-weighted MR data were first bias corrected using N4ITK bias field correction~\cite{Tustison2010}. Each individual’s CT image was rigidly realigned and resampled to their T1-weighted image using NiftyReg~\cite{Modat2014} with cubic spline interpolation, after the scanner bed was edited out of the CT images using custom scripts. In the PET Centre dataset, where voxel sizes were $\leq$ 1.1 mm, we did not apply any resampling but instead allowed the network treat this as a variation in head size, similar to a scaling augmentation (± 10\%). In the clinical paediatric dataset, we resampled all data to 1 mm\textsuperscript{3} isotropic resolution to make it compatible with the resolution of the PET Centre dataset. 
\subsection{Network architecture}
\begin{figure}[ht]
{\centering\includegraphics[width=0.65\textwidth]{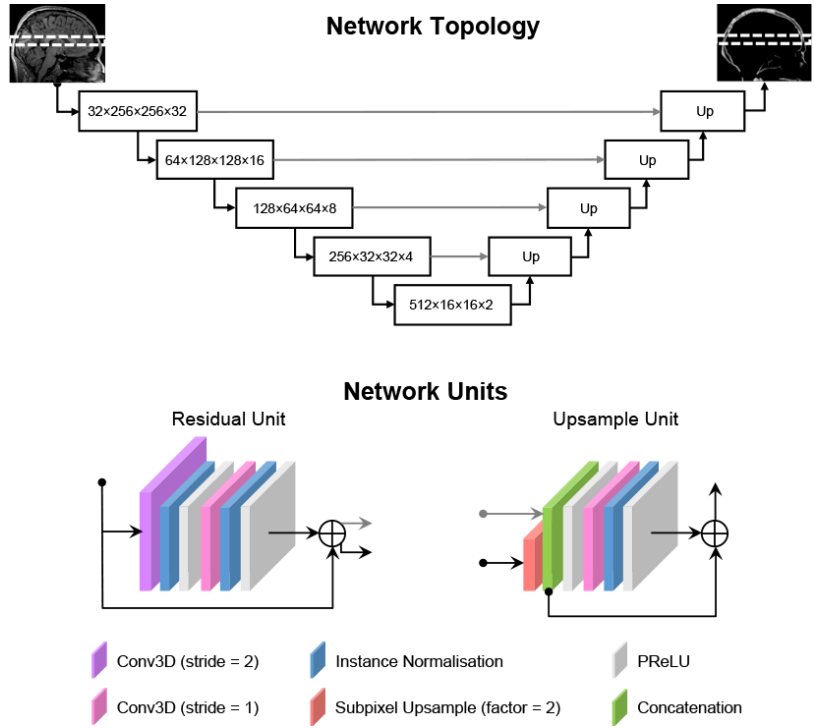}
\caption{3D residual U-Net with subpixel upsampling. Overall network topology (top panel) and details of network units (bottom panel). In the top panel, the number of channels and tensor size are shown in each layer of the encoding path. Conv3D: 3D convolutions;  PReLU: Parametric Rectified Linear Unit.} \label{fig1}}
\end{figure}

The convolutional neural network (CNN) we used for image translation is based on a residual U-Net architecture~\cite{Ronneberger2015,He2015,Kerfoot2019}. The network consists of blocks of 3D convolutions (kernel size = 3), each followed by instance normalisation~\cite{Ioffe2015} and a Parametric Rectified Linear Unit (PReLU)~\cite{He2015a} activation. Our U-Net has five layers of these convolutional blocks with 32, 64, 128, 256 and 512 channels in each consecutive layer. In the encoding path, instead of max pooling, downsampling is performed using convolutions with a stride of 2 and a residual connection is added. In the decoding path, we use a modified 3D version of the pixel shuffle method~\cite{Shi2016} to upsample data from lower layers by a factor of 2. The upsampled input is then concatenated with the output from the encoding path in the same layer. Figure~\ref{fig1} illustrates the overall topology of our U-Net and the network units in the encoding and decoding paths. 
\subsection{Network implementation}
We implemented the above network architecture in PyTorch \url{https://pytorch.org} using the MONAI framework~\cite{Cardoso2022} \url{https://monai.io} on an NVIDIA Quadro M4000 GPU with 8GB of RAM. Our network took a 3D patch from the T1-weighted MR as input and produced the corresponding CT patch. 

In order to normalise intensities across each dataset, histogram normalisation was applied to the input T1-weighted MR images using the \textit{HistogramNormalize} transform in MONAI. MR images were then rescaled between -1 and 1 (\textit{ScaleIntensity}) and CT images were thresholded at -1000 Hounsfield units (HU) and clipped above 2400 HU (\textit{ThresholdIntensity}). 

The network was trained on batches of eight patches each of size 256 × 256 × 32 voxels sampled randomly from the pairs of images. We used the Adam optimiser with weight decay regularisation~\cite{Loshchilov2019} with an initial learning rate of 3 × 10\textsuperscript{-3}, $\beta$\textsubscript{1} = 0.95, $\beta$\textsubscript{2} = 0.99, and weight decay = 0.03 (hyperparameters chosen based on~\cite{Gugger2018}). We defined an epoch as one iteration of a randomly sampled patch over all images in the training dataset. 

We performed a five-fold cross-validation of the network, dividing the data into 80\% training and 20\% testing data for each fold, with one image set from the testing data used for validation. We used the mean squared error loss and tracked the mean absolute error of the validation dataset every five epochs. The network was trained until the validation loss did not improve by more than 1 HU over 100 epochs (approximately 900--1000 epochs in total for all folds). At this point, the learning rate was reduced to 3 × 10\textsuperscript{-4} and the network further trained until the validation loss did not improve by more than 1 HU over 100 epochs (approximately 100 to 450 epochs). We took the network weights at the epoch showing the best validation loss as the final weights for each fold of the cross-validation. The full 3D pseudo-CT volumes were produced from T1-w MR images of each test dataset by predicting overlapping patches (\textit{sliding\_window\_inference}, overlap = 0.9, blend mode =  gaussian). 
\subsection{Evaluation of pseudo-CT accuracy}
We evaluated the accuracy of pseudo-CTs produced using our deep learning method against reference CTs using the mean absolute error (MAE) metric~\cite{Burgos2014} defined as MAE=$\Sigma$($|pCT-CT|$/V), where V is the number of non-zero voxels within a head mask. The head mask was obtained by cropping the T1-weighted MR image to have a 17cm cranio-caudal extent so as to exclude the neck using \textit{robustfov} (\url{https://fsl.fmrib.ox.ac.uk}). We also evaluated the Jaccard coefficient (JC;~\cite{Jaccard1901}) of overlap within a skull mask, which was obtained by thresholding the CT image within the head mask at 300 HU to obtain intensities relating to the skull.
\subsection{Comparison with MaxProb method}
The MaxProb method~\cite{Merida2017} was implemented on a Linux machine at the KCL \& GSTT PET Centre to produce linear attenuation maps (µ-maps) for all 110 individuals. For the comparison with µ-maps produced by the MaxProb method, we first converted the reference CTs and pseudo-CTs produced by our method to 511 keV µ-map according to the bilinear equation below~\cite{Carney2006}:
\\Below Break Point: $\mu = 9.6\times 10^{-5} (HU + 1000)$ cm\textsuperscript{-1},
\\Above Break Point: $\mu = a (HU + 1000) + b$ cm\textsuperscript{-1},
\\with the Break Point at 1030 HU, a = 5.64 × 10$^{-5}$ cm\textsuperscript{-1} and b = 4.08 × 10$^{-2}$ cm\textsuperscript{-1}, based on a 140 kVp CT tube voltage.

We evaluated the accuracy of the µ-maps obtained from our pseudo-CTs (µ\textsubscript{pCT}) to µ-maps from the reference CTs (µ\textsubscript{CT}) using the relative MAE (rMAE) metric~\cite{Burgos2014} defined as $rMAE_{pCT} = 100 \times \Sigma(|\mu _{pCT}-\mu _{CT}|)/\Sigma \mu _{CT}$. We also evaluated rMAE on the µ-maps obtained from the MaxProb method. We compared rMAE from our deep learning method against those obtained from the MaxProb method with a paired t-test.
\subsection{Transfer learning in paediatric dataset}
We combined our deep learning model by redoing the training on the full set of images in the PET Centre dataset, using the same training scheme as in the cross-validation. We first implemented our fully trained network on the clinical paediatric dataset without refinement. Next, we investigated whether we could improve the performance of our network in paediatric data through transfer learning. Using the fully trained network as a starting point, we further trained the network on the clinical paediatric dataset only using the same training scheme as described above. We performed another 4-fold cross-validation in the clinical paediatric dataset and evaluated the resulting pseudo-CT against reference CT in each fold as above.
\section{Results}
\subsection{Pseudo-CT accuracy}
\begin{figure}[ht]
{\centering\includegraphics[width=0.98\linewidth]{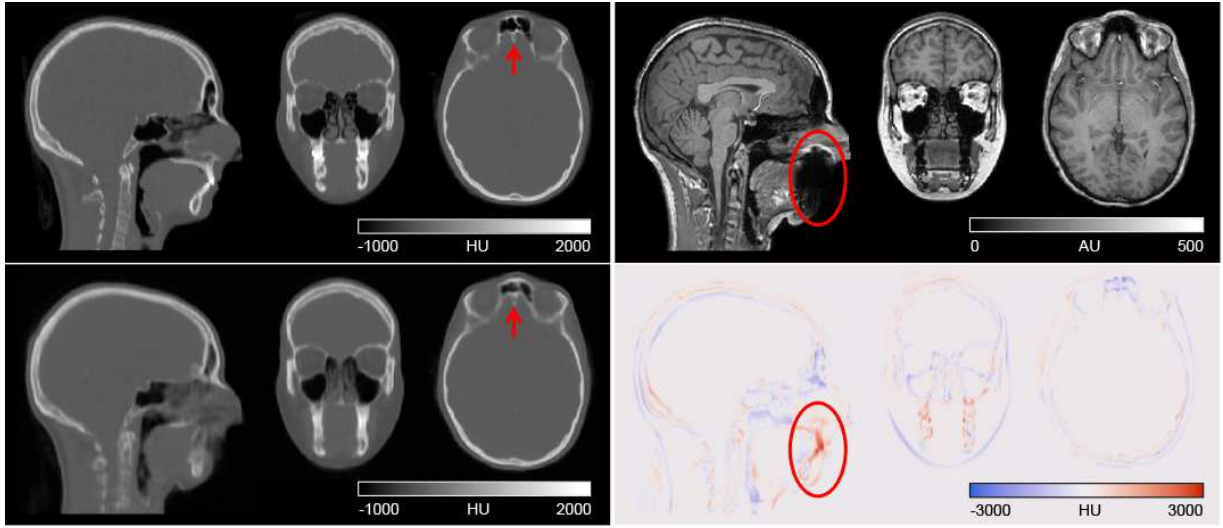}
\caption{
Reference CT (top left) and corresponding T1-weighted MR image (top right) for a representative individual shown with the pseudo CT (bottom left) generated using our network and the difference image (reference--pseudo CT; bottom right). The red arrows show fine bone structures, which are reproduced in the pseudo CT. The region with the largest error is circled in red and corresponds to MR signal dropout in the mouth due to dental fillings.\\
}\label{fig2}}
\end{figure}

Figure \ref{fig2} shows the reference CT and pseudo-CT produced by our deep learning method for a representative individual, together with the difference image (reference CT–pseudo-CT) and T1-weighted MR used to generate the pseudo-CT. Qualitatively, our pseudo-CT looks similar to the ground truth CT and our method was able to faithfully reproduce fine bone structures, for example in the tissue and air-space near the frontal sinuses (red arrow in Figure 2). The area with the largest error coincided with signal dropout on the MR image due to dental fillings (circled in Figure 2). On average, our method produced pseudo-CT with MAE of 100.4 ± 21.3 HU compared to reference CT within the head mask. The Jaccard coefficient (JC) for the amount of overlap between the skull masks from our pseudo-CT and the reference CT was 0.73 ± 0.07. Table \ref{table2} shows the results of each fold of the cross-validation.
\begin{table}[ht]
\centering
\caption{Accuracy of pseudo-CTs compared with reference CTs in cross-validation.}\label{table2}
\begin{tabularx}{0.4\textwidth}{ l *{2}{|Y} }
\hline
\multicolumn{1}{c|}{Fold} & \textbf{MAE (HU)} & \textbf{JC}\\
\hline
\multicolumn{1}{c|}{1} & 104.1 ± 28.4 & 0.73 ± 0.05\\
\multicolumn{1}{c|}{2} & 93.1 ± 9.5 & 0.75 ± 0.05\\
\multicolumn{1}{c|}{3} & 102.5 ± 19.6 & 0.71 ± 0.07\\
\multicolumn{1}{c|}{4} & 96.2 ± 16.7 & 0.75 ± 0.05\\
\multicolumn{1}{c|}{5} & 106.0 ± 25.1 & 0.72 ± 0.09\\
\hline
\multicolumn{1}{c|}{\textbf{Mean}} & \textbf{100.4 ± 21.3} & \textbf{0.73 ± 0.07}\\
\hline
\multicolumn{3}{l}{Values are given as mean ± SD.}
\end{tabularx}
\end{table}
\subsection{Comparison with MaxProb method}
\begin{figure}[ht]
{\centering\includegraphics[width=0.98\linewidth]{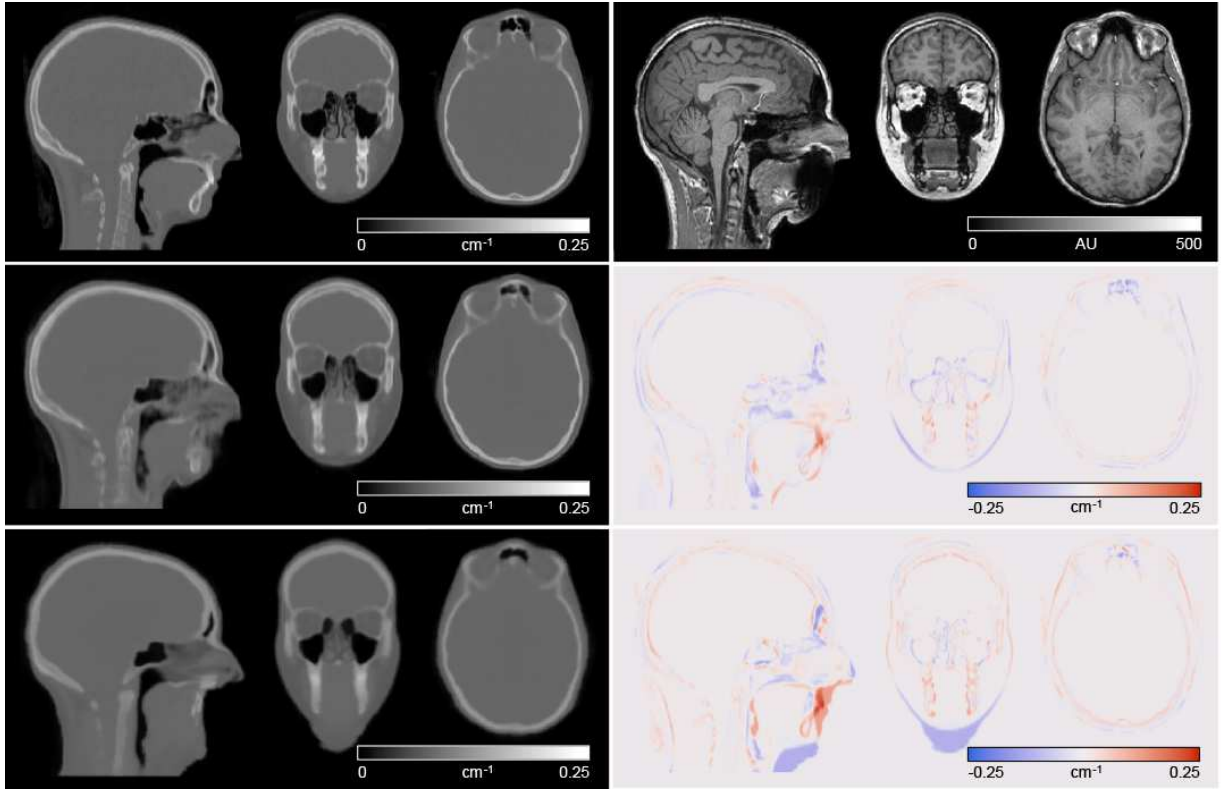}
\caption{
Top row: µ-map from reference CT (left) and corresponding T1-weighted MR image (right) for the same individual as in Figure 2. Middle row: µ-map from our pseudo CT method (left) and corresponding difference image (reference--pseudo-CT); right). Bottom row: µ-map from MaxProb method (left) and difference image (right).
} \label{fig3}}
\end{figure}

Figure \ref{fig3} shows the µ-maps derived from reference CT, the MaxProb method and our pseudo-CT along with difference images, for the same individual as in Figure 2. Our method was able to learn to generate some intensities in the area where there was signal dropout on the MR image due to dental fillings, whereas the MaxProb method was not able to. The fine bone structures in the air space near the frontal sinuses were not reproduced with the MaxProb method. Compared with reference CT µ-maps, our network produced µ-maps with smaller errors than the MaxProb method: mean rMAE was 8.4 ± 2.1\% for our pseudo-CT µ-maps and 13.1 ± 1.5\% for the MaxProb µ-maps. 
\subsection{Implementation of MR-AC in paediatric dataset and impact of pre-training}
\begin{figure}[ht]
{\centering\includegraphics[width=0.98\linewidth]{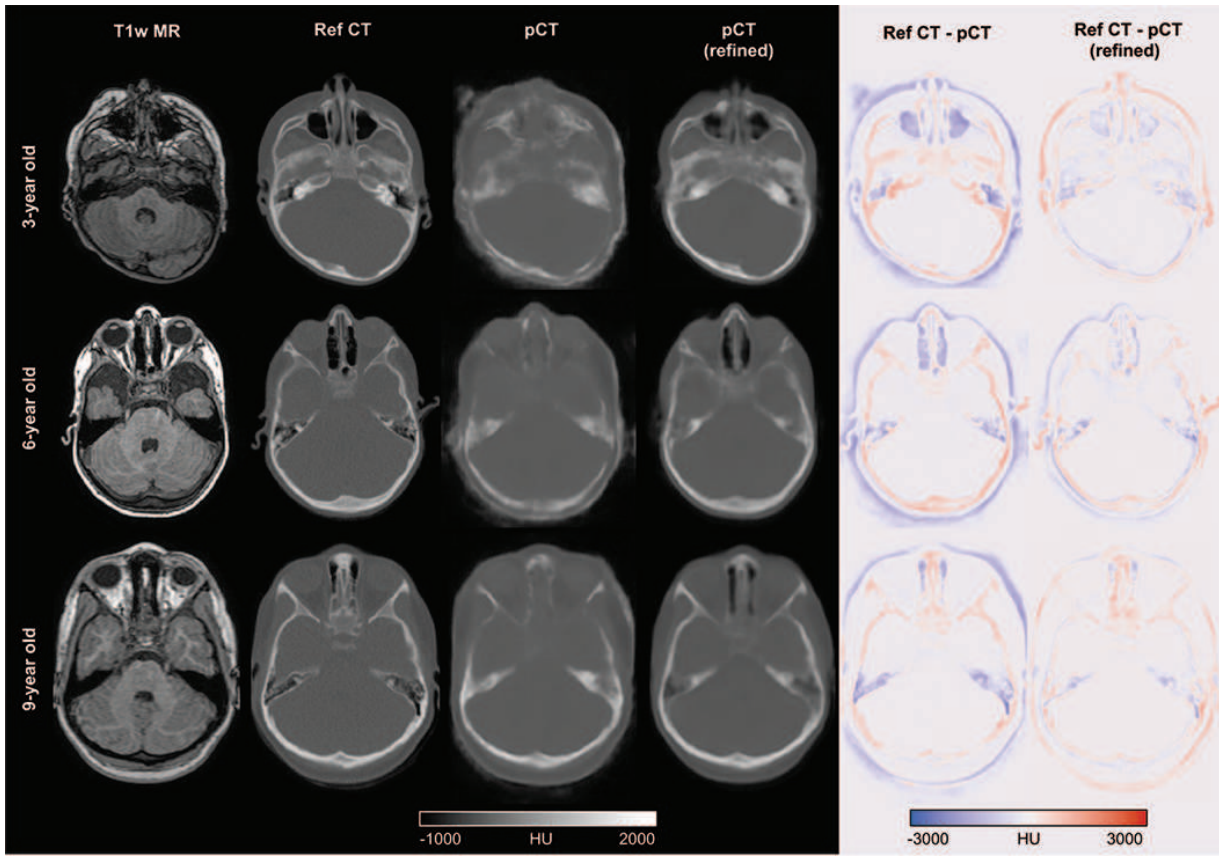}
\caption{
Example images from a 3-year-old (top row), 6-year-old (middle row) and 9-year-old (bottom row) child from the clinical paediatric dataset. Images are (from left to right): T1-weighted MR, reference CT, pseudo CT generated using the network trained on PET Centre data, pseudo CT generated after refinement in paediatric data, difference image between reference CT and pseudo CT without and with refinement.
} \label{fig4}}
\end{figure}

Figure \ref{fig4}  shows the T1-weighted MR, corresponding reference CT, pseudo-CTs generated using the network as trained on adult data (i.e. without refinement), and pseudo-CTs generated using the network refined on paediatric data, together with difference images (reference CT–pseudo-CT) for three randomly selected individuals aged 3, 6 and 9 years. Qualitative inspection shows that the pseudo-CT (without refinement) generated in the 3-year-old shows the highest errors due to thinner skull structures not being reproduced. This was slightly better in the 6- and 9-year-old patients. Pseudo-CTs generated using the refined network were qualitatively better in all three cases. MAE improved from 174.7 ± 33.6 HU when using the existing network to 127.3 ± 39.9 HU after transfer learning in the paediatric dataset.
\section{Discussion}
Our proposed method is able to generate a pseudo-CT with comparable accuracy to existing deep learning methods trained in adults. The µ-maps produced using our method have higher accuracy than the MaxProb multi-atlas method when compared with reference CT-based µ-maps. We refined our network on a clinical paediatric dataset and showed the benefit of transfer learning for improving the accuracy of pseudo-CT in paediatric data. The results suggest that the network can be used to produce pseudo-CTs in paediatric datasets for use in PET-MR-AC.

Our method produces pseudo-CT with MAE (100.4 ± 21.3) and Jaccard coefficients of overlap in skull (0.73 ± 0.07) comparable to other published deep learning methods (c.f. Jaccard coefficient in bone tissue of 0.70 in~\cite{Ladefoged2019}). The method outperforms a well-performing multi-atlas method in terms of µ-map accuracy. In the PET Centre dataset, all five folds of the cross-validation showed similar pseudo-CT accuracy metrics, demonstrating the robustness of the network and training scheme, and balance of the training dataset. 

We show proof of concept that a deep learning method trained on predominantly adult data can be refined and implemented in paediatric data, and showed the added benefit of transfer learning on clinical data where skull geometry and size may differ from the training dataset. Further work will include validation of the PET-MR AC method in both datasets.
\subsubsection*{Acknowledgements.}
S.N.Y. is supported by the MRC grant MR/T023007/1. This work was supported by the UK Department of Health via the NIHR Comprehensive Biomedical Research Centre Award (COV-LT-0009) to Guy's and St Thomas' NHS Foundation Trust (in partnership with King's College London and King's College Hospital NHS Foundation Trust), and by the Wellcome Engineering and Physical Sciences Research Council Centre for Medical Engineering at King's College London (WT 203148/Z/16/Z).
%
\bibliographystyle{splncs04}
\bibliography{ms.bib}
\end{document}